\documentclass[prl,twocolumn,showpacs,showkeys,preprintnumbers,aps,amssymb]{revtex4}

\usepackage{amsmath,amssymb}
\usepackage{graphicx}
\usepackage{dcolumn}
\usepackage{bm}
\newcommand{\Exp}[1]{\textrm{e}^{#1}}
\begin{document}

\preprint{KEK-TH-966}

\title{Two-pion bound state in sigma channel at finite temperature}

\author{Yoshimasa HIDAKA}
 \email{hidaka@post.kek.jp}

\author{Osamu MORIMATSU}
 \email{osamu.morimatsu@kek.jp}
\author{Tetsuo NISHIKAWA}
 \email{nishi@post.kek.jp} 
\affiliation{%
Institute of Particle and Nuclear Studies, High Energy Accelerator Research Organization, 1-1, Ooho, Tsukuba, Ibaraki, 305-0801, Japan
}
\author{Munehisa OHTANI}
\email{ohtani@rarfaxp.riken.jp} 
\affiliation{%
Radiation Laboratory, RIKEN, Wako, Saitama, 351-8902, Japan
}

\date{\today}

\begin{abstract}
We study how we can understand the change of the spectral function and the pole location of the correlation function for sigma at finite temperature, which were previously obtained in the linear sigma model with a resummation technique called optimized perturbation theory.
There are two relevant poles in the sigma channel.
One pole is the original sigma pole which shows up as a broad peak at zero temperature and becomes lighter as the temperature increases.
The behavior is understood from the decreasing of the sigma condensate, which is consistent with the Brown-Rho scaling. 
The other pole changes from a virtual state to a bound state of $\pi\pi$ as the temperature increases which causes the enhancement at the $\pi\pi$ threshold.
The behavior is understood as the emergence of the $\pi\pi$ bound state due to the enhancement of the $\pi\pi$ attraction by the induced emission in medium.
The latter pole, not the former, eventually degenerates with pion above the critical temperature of the chiral transition.
This means that the observable \lq\lq$\sigma$" changes from the former to the latter pole, which can be interpreted as the level crossing of $\sigma$ and $\pi\pi$ at finite temperature.
\end{abstract}
\maketitle

Restoration of chiral symmetry at finite temperature is 
an interesting topic in hadron physics
and recently discussed in connection
with deconfinement transition \cite{Fukushima:2003fm,
Hatta:2003ga, Digal:wn} as well as
mixed chiral condensate in lattice simulation \cite{Doi:2003gh}.
Even below the critical temperature $T_c$, 
drastic change of the meson spectrum is expected as a precursor 
for the restoration of chiral symmetry \cite{Hatsuda:eb},
and actually strong enhancement of $\sigma$ spectrum near 
the $\pi\pi$ threshold was reported \cite{Chiku:1997va}.
This spectrum enhancement is naturally understood based on
the partial restoration of chiral symmetry as follows;
As temperature increases,
$\sigma$ diminishes its mass while $\pi$ becomes heavier
because their masses degenerate above $T_c$.
This demands that the mass of $\sigma$ coincides with 
twice that of $\pi$ at certain temperature and the phase space of the decay $\sigma\rightarrow\pi\pi$ is squeezed.
This squeeze causes the spectrum enhancement.  
In view of the above we analyzed complex poles of the propagator and elucidated that
two poles are dominant on the $\sigma$ spectrum \cite{Hidaka:2003mm}.
One of these poles corresponds to the broad bump at $T=0$ and the other 
pole causes the threshold enhancement.

In this paper we study how the behavior of these poles is understood.
We show that the behavior of the pole, which causes the threshold enhancement, is understood as the emergence of the $\pi\pi$ bound state due to the enhancement of the $\pi\pi$ attraction by the induced emission in medium.

Let us briefly review the results of \cite{Hidaka:2003mm}, in which we calculated the spectral function and pole locations using the ${\cal O}(4)$ linear sigma model with a resummation technique called the optimized perturbation theory (OPT) \cite{Chiku:1997va,Stevenson:1981vj}. 
The lagrangian of the ${\cal O}(4)$ linear sigma model is as follows:
\begin{equation}
  {\cal L}= \frac{1}{2}(\partial\phi_{\alpha})^2-\frac{1}{2}\mu^2 
  \phi_{\alpha}^{2}-
             \frac{\lambda}{4!}(\phi_{\alpha}^{2})^2 +h\phi_0,
\end{equation}
with $\phi_\alpha=(\phi_0,\bm{\pi})$. 
In order to describe the spontaneously broken chiral symmetry at low temperature ($T$), $\mu^2$ must be negative.
Then,  the field $\phi_0$ has a non-vanishing expectation value $\xi$,
which makes us decompose the field operator $\phi_0$
into the classical condensate and the quantum fluctuation as
$\phi_0 = \xi + \sigma\;$.  
The renormalized parameters are determined by the experimental values of the pion mass ($m_{\pi}$), the pion decay constant ($f_\pi$) and the peak energy of the sigma spectral function
$\rho_\sigma$ at $T=0$ following \cite{Chiku:1997va}:
$\mu^2=-(283\,{\textrm{ MeV}})^2,\ \lambda=73.0, \ h=(123\,{\textrm{ MeV}})^3$.

The one-loop effective potential is given by
\begin{eqnarray}
 V^{\rm eff}(\xi)&=&-h\xi+\frac{\mu^2}{2}\xi^2+\frac{\lambda}{4!}\xi^4
   +\frac{1}{2}T\sum\int\frac{\textrm{d}^3p}{(2\pi)^3}\nonumber\\
   &&\times [\ln(p^2+m_{0\sigma}^2)+3\ln(p^2+m_{0\pi}^2)],
\end{eqnarray}
where $\sum$ denotes a summation of the Matsubara frequency.
$m_{0\sigma}$ and $m_{0\pi}$ are the tree-level masses of 
the fluctuations defined by
\begin{equation}
m_{0\sigma}^2=m^2+\lambda \frac{\xi^2}{2} \; ,
 \ \ m_{0\pi}^2=m^2+\lambda \frac{\xi^2}{6} \; .
\label{eq:treemass}
\end{equation}
Here, we have already introduced the OPT and $m(T)$ is an optimal mass parameter, which will be explained shortly.
As a minimum of the effective potential, the condensate $\xi(T)$ 
is determined by the stationary condition,
\begin{equation}
  0=\frac{\partial V^{\textrm{eff}}}{\partial \xi} =-h+\mu^2\xi
   +\frac{\lambda}{3!}\xi^3 
   +\frac{\lambda \xi}{2}(I^{(1)}_\sigma +I^{(1)}_\pi).
\label{eq:gap}
\end{equation}
The last term is the contribution from the tadpole diagrams, where we adopt the modified minimal subtraction scheme ($\overline{\rm MS}$):
\begin{equation}
  I^{(1)}_{\phi} 
       \stackrel{\overline{\rm MS}}{=} -\frac{m_{0\phi}^2}{16\pi^2}(
        1-\ln\frac{m_{0\phi}^2}{\kappa^2}) 
 +\int_0^\infty\frac{{\rm d}p\,p^2}{2\pi^2}
 \frac{n_\phi}{\omega_\phi}\;,
  \label{eq:I1}
\end{equation}
with $\omega_\phi=\sqrt{\bm{p}^2+m_{0\phi}^2}$,
$n_\phi=({\rm e}^{\omega_\phi/T}-1)^{-1}$ and
$\kappa$ is the renormalization point: $\kappa=255$MeV.

The idea of the OPT is to incorporate as much thermal effects as possible into the optimal parameter(s)  \cite{Stevenson:1981vj}.
The original mass term $\mu^2$ is decomposed into the optimal mass parameter $m^2(T)$ and the rest $\chi\equiv m^2(T)-\mu^2$.
The mass term with $m^2(T)$ is treated as a nonperturbative part while the term with $\chi$ as a perturbation.
We determine $m(T)$ by the condition that (the thermal part of) the one-loop correction vanishes for the pion mass at zero momentum \cite{Chiku:1997va}.

\begin{figure}[t]
    \includegraphics[width=1.\linewidth]{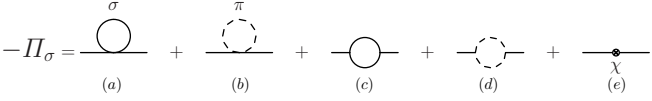}\\
    \caption{Feynman diagrams representing one-loop self-energy for $\sigma$. Solid and dashes lines correspond to $\sigma$ and $\pi$, respectively.}
    \label{self-energy}
\end{figure}

The spectral function, which is the imaginary part of 
the propagator, is related to the self-energy $\varPi_\sigma(k,T)$ 
as
\begin{eqnarray}
  \label{eq:spec}
\rho_\sigma(k,T)&=& 2\ {\rm Im}\ D_\sigma(k,T)\cr
&=&
 -2 \ {\rm Im}\ (k^2-m_{0\sigma}^2-\varPi_\sigma(k,T))^{-1},
\end{eqnarray}
with $k=(\omega,\boldsymbol{k})$. 
At one-loop level, the self-energy shown in Fig.\ref{self-energy} is written as
\begin{eqnarray}
  \varPi_\sigma(k,T) &=& -(m^2-\mu^2)+\frac{\lambda}{2}\left(I^{(1)}_\sigma
    +I^{(1)}_\pi\right)\cr
    &&-\frac{\lambda^2\xi^2}{2}\left(
 I^{(2)}_\sigma+\frac{1}{3}I^{(2)}_\pi\right) ,
  \label{eq:selfen}
\end{eqnarray}
where $I^{(2)}_\phi$ corresponds to a bubble diagram:
\begin{eqnarray}
I^{(2)}_{\phi} &\stackrel{\overline{\rm MS}}{=}&
   \frac{1}{16\pi^2}\left(2-\ln\frac{m_{0\phi}^2}{\kappa^2}
    +c\,\ln \frac{c-1}{c+1}   \right) \nonumber
   \\ && \hspace{3em} 
 -\frac{1}{\pi^2}\int_0^\infty\frac{{\rm d}p \,p^2}{\omega_\phi}
   \frac{n_\phi}{\omega^2-4\omega_\phi^2}\; ,
  \label{eq:I2}
\end{eqnarray}
with $c=\sqrt{1-4m_{0\phi}^2/k^2}$ .

It is useful to know the temperature dependence of the pole locations of $D_{\sigma}(k)$ in order to understand the behavior of the spectral function.
This is because the spectrum is generally dominated by the poles of the propagator, and hence we locate poles of $D_{\sigma}(k)$.
For simplicity the spatial momentum is fixed to $\bm{k}=0$, but the generalization to $\bm{k}\neq0$ is not difficult \cite{Hidaka:2002xv}.
The inverse of the propagator $D^{-1}_{\sigma}(k)$ is written as
\begin{equation}
D^{-1}_{\sigma}(k)=\omega^2-m_{0\sigma}^2-\varPi_{\sigma}(k,T).
\end{equation}
The position of the pole is determined by the equation $D^{-1}_{\sigma}(k)=0$ on the complex energy plane.
The inverse propagator, $D^{-1}_{\sigma}(k)$, is analytic on the complex plane except for the real axis and has branch cuts for $|\omega| > 2m_{0\pi}$ and $|\omega| > 2m_{0\sigma}$ at one loop level.
Accordingly, we perform the analytic continuation of $D_\sigma^{-1}$
following \cite{Patkos:2002vr,Hidaka:2002xv} to locate the relevant poles. Through different cuts one goes to different Riemann sheets on which different analytic continuations are defined.
We are interested in the Riemann sheet analytically continued through the cut $\omega > 2m_{0\pi}$ from the upper half plane, which we call the second Riemann sheet.
We define the inverse of the propagator on the second Riemann sheet as 
\begin{equation}
D^{-1}_{\sigma,\textrm{2nd}}(k)=D^{-1}_{\sigma,\textrm{1st}}(k)+F(k).
\end{equation}
Here $D^{-1}_{\sigma,\textrm{1st}}(k)$ is the inverse of the propagator on the first Riemann sheet and $F(k)$ is determined by the condition that $D^{-1}_{\sigma,\textrm{2nd}}(k)$ is continuous to $D^{-1}_{\sigma,\textrm{1st}}(k)$ on the real axis $2m_{0\pi}<\omega<2m_{0\sigma}$. 
Therefore $F(k)$ is given by discontinuity for $2m_{0\pi}<\omega<2m_{0\sigma}$:
\begin{eqnarray}
F(k)&=&\textrm{Disc} D^{-1}_{\sigma}(k)\nonumber\\
    &=&\frac{(\lambda\xi)^2}{6}\frac{i}{8\pi}\Bigl(1-\frac{4m_{0\pi}^2}{\omega^2}\Bigr)^{\frac{1}{2}}\Bigl(1+2\frac{1}{\Exp{\frac{\beta \omega}{2}}-1}\Bigr).
\end{eqnarray}
$F(k)$ is real for $0<\omega<2m_{0\pi}$ on the second Riemann sheet:
\begin{equation}
F(k)=\frac{(\lambda\xi)^2}{6}\frac{1}{8\pi}\Bigl(\frac{4m_{0\pi}^2}{\omega^2}-1\Bigr)^{\frac{1}{2}}\Bigl(1+2\frac{1}{\Exp{\frac{\beta \omega}{2}}-1}\Bigr).
\end{equation}
Thus, $D^{-1}_{\sigma,\textrm{2nd}}(k)$ is also real for $0<\omega<2m_{0\pi}$.
From the above we can show without explicit calculation that when the propagator has no pole on the first Riemann sheet the propagator has a pole in the interval $0<\omega<2m_{0\pi}$ on the second Riemann sheet as follows.
Firstly, the stability of the ground state demands 
that $D^{-1}_{\sigma \textrm{1st}}$ is negative at $\omega = 0$. Here we note
that $D^{-1}_{\sigma \textrm{1st}}$ is real and has no pole on the real axis 
below the $\pi\pi$ threshold. It follows that 
$D^{-1}_{\sigma \textrm{1st}}$ is negative also at $\omega =2m_{0\pi}$,
which in turn assures us of negative $D^{-1}_{\sigma \textrm{2nd}}$ at this point. 
Secondly, we see that $D^{-1}_{\sigma \rm 2nd} \to \infty$ as $\omega\rightarrow 0$ because $F(k)\rightarrow +\infty$ as $\omega\rightarrow 0$.
From these facts, $D_{\sigma \textrm{2nd}}$ necessarily has at least a pole on the real axis below the $\pi\pi$ threshold.
The number of poles, however, cannot be determined from the above argument.
\begin{figure}[t]
    \includegraphics[width=1.\linewidth]{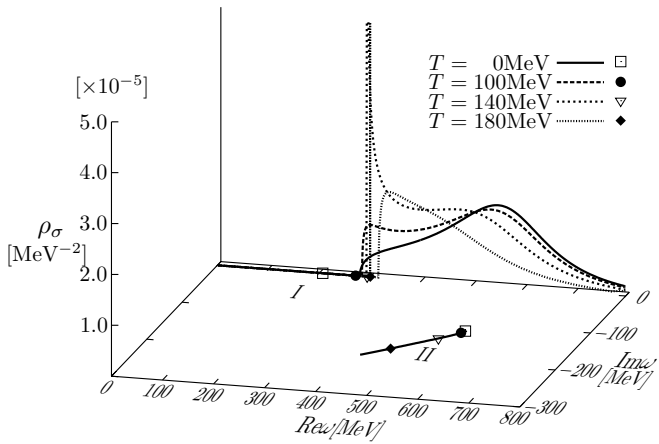}\\
    \caption{Spectral function and the pole position at $T=0,100, 140$ and $180$ MeV for $\bm{k}=0$ in the $\sigma$ channel.}
    \label{spectral3D}
\end{figure}

In fact, we numerically confirm only one pole on the real axis of the 2nd Riemann sheet in the low temperature region as expected.
And we also find another pole on the 2nd Riemann sheet.
Here we show the results of explicit calculation for the spectral functions and the pole locations of the propagators at finite temperature with the spatial momentum kept fixed ($\bm{k}=0$).
Figure \ref{spectral3D} shows the spectral function and the poles of the propagator for sigma at finite temperature which have been obtained in \cite{Chiku:1997va,Hidaka:2002xv}.
There are two relevant poles which we call Pole {\sc I} and Pole {\sc II} on the complex energy plane.
Pole {\sc II} corresponds to a broad bump around $\omega=550$MeV at $T=0$.
Pole {\sc II} moves left and down as $T$ increases.
The peak of the spectral function also shifts left and becomes broader.
As discussed above, there exists Pole {\sc I} on the real axis on the second Riemann sheet.
A little shoulder of the spectral function at the $\pi\pi$ threshold seems to reflect the effect of Pole {\sc I}.
As $T$ increases, Pole {\sc I} moves right on the real axis and approaches the $\pi\pi$ threshold,
which causes the enhancement at the $\pi\pi$ threshold.
When $T=145$ MeV, Pole {\sc I} crosses the $\pi\pi$ threshold and appears on the first sheet.
The results imply that the observable \lq\lq$\sigma$" changes from Pole 
{\sc II} to Pole {\sc I} as $T$ increases and at certain 
temperature($T\approx
120$MeV) spectrum with twin peaks is observed owing to two poles.
\begin{figure}[b]
    \includegraphics[width=1.\linewidth]{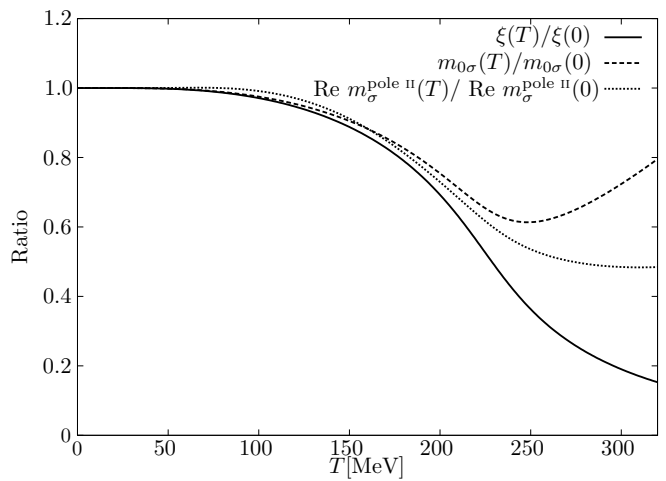}\\
    \caption{Ratio of the condensate $\xi$, the tree level mass $m_{0\sigma}$ and the real part of the pole {\sc II} Re$m_{\sigma}^\textrm{pole {\sc II}}$ between  $T\neq0$ and $T=0$.}
    \label{scaling}
\end{figure}

Now, we discuss how the behavior of poles can be understood.
In figure \ref{scaling}, we plot as a function of the temperature the condensate $\xi$, the tree level mass $m_{0\sigma}$ and the real part of Pole {\sc II} Re$m_{\sigma}^\textrm{pole \sc{II}}$ normalized by their values at zero temperature.
Below $T_c(\approx200$MeV), one sees the following scaling:
\begin{equation}
\frac{\xi(T)}{\xi(0)}\approx\frac{m_{0\sigma}(T)}{m_{0\sigma}(0)}\approx\frac{m_{\sigma}^\textrm{pole {\sc II}}(T)}{m^\textrm{pole {\sc II}}_{\sigma}(0)}.
\end{equation}
This is consistent with the Brown-Rho scaling \cite{Brown:kk}.
Accordingly, the behavior of Pole {\sc II} can be understood from decreasing the sigma condensate.
On the other hand, the behavior of Pole {\sc I} is understood as the generation of a bound state of two pions as follows.
The pion loop diagram Figure.\ref{self-energy} (d) is dominant for the spectral function of the sigma around the $\pi\pi$ threshold. 
The contribution of the pion loop diagram at $T=0$ with the spatial momentum kept fixed ($\bm{k}=0$) is as follows,
\begin{eqnarray}
\varPi^{(\pi\pi)}_{\sigma}(\omega,T=0)&=& -\frac{(\lambda\xi)^2}{6}I^{(2)}_{\pi}(k,T=0)\nonumber\\
&=&\frac{(\lambda\xi)^2}{6}\int\frac{\textrm{d}^3p}{(2\pi)^3}\frac{1}{\omega_\pi}\frac{1}{\omega^2-4\omega_\pi^2}.
\end{eqnarray}
This process gives attraction between two pions below the $\pi\pi$ threshold ($\omega<2m_{0\pi}$) because the integrand is negative.
If the coupling is sufficiently strong, this attraction makes two pions bound.
We then show that in the heat bath the effective coupling becomes actually stronger and a bound state is generated.
In the heat bath the coupling in the self energy is changed by the induced emission as
\begin{eqnarray}
&&\frac{(\lambda\xi(T=0))^2}{6}\nonumber\\
&\rightarrow& \frac{(\lambda\xi(T))^2}{6}\Big((1+n_\pi)(1+n_\pi)-n_\pi n_\pi\Big)\nonumber\\
&=&\frac{(\lambda\xi(T))^2}{6}(1+2n_\pi).
\end{eqnarray}
Here $n_\pi$ is the Bose-Einstein distribution function and 
the statistical weight $(1+n_\pi)(1+n_\pi)$ is for the process $\sigma\rightarrow\pi\pi$ and $n_\pi n_\pi$ for the inverse process $\pi\pi\rightarrow\sigma$.
Accordingly, at finite temperature the sigma self-energy becomes 
\begin{equation}
\varPi^{(\pi\pi)}_{\sigma}(\omega,T)=\frac{1}{6}\int\frac{\textrm{d}^3p}{(2\pi)^3}\frac{1}{\omega_\pi}\frac{(\lambda\xi(T))^2(1+2n_\pi)}{\omega^2-4\omega_\pi^2}.
\end{equation}
Clearly, the induced emission makes the coupling effectively stronger by factor $1+2n_\pi$.
However,  $\xi(T)$ becomes smaller as $T$ increases which makes the coupling effectively weaker.
Therefore, whether the effective coupling becomes stronger or weaker is determined by the competition of the above two factors.
If the effect of the induced emission is dominant, a bound state of two pions shows up as a pole on the first Riemann sheet.
Otherwise the original sigma pole (Pole {\sc II}), which corresponds to a broad bump at $T=0$, would move to the first Riemann sheet.
In our case, $\xi(T)$ does not change much except near the critical temperature, therefore the former situation is realized.

The latter possibility was discussed in ref. \cite{Patkos:2002vr}, where in the leading  
order of the large-$N$ approximation it was obtained that the sigma pole at $T=0$
continuously moves on the first Riemann sheet by passing through a virtual
state located on the real axis of the second Riemann sheet.
The above explanation of the behavior of the poles does not seem to apply to
the result of ref.\cite{Patkos:2002vr}, in which below a certain value 
of $T$ there is no pole corresponding to Pole 
{\sc I} and the vacuum expectation value also does not vary too much with the temperature.
In ref.\cite{Patkos:2002vr} the trajectory of the sigma pole at $T=0$ seems to be a consequence of the
resummation of the pion bubble which changes the form of the sigma propagator
with respect to the one obtained in a one-loop analysis.
\begin{figure}[t]
    \includegraphics[width=1.\linewidth]{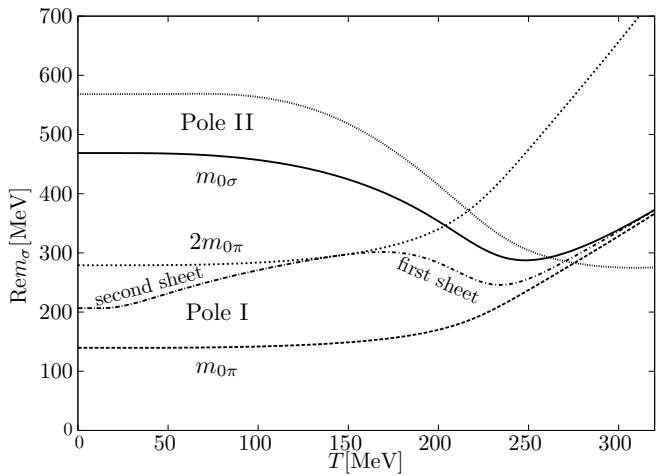}\\
    \caption{Real part of complex poles of $D_\sigma$, tree level masses $m_{0\sigma}$, $m_{0\pi}$ and $2m_{0\pi}$.}
    \label{mass2}
\end{figure}

In figure \ref{mass2}, we plot the real part of the complex poles of $D_\sigma$, tree level masses $m_{0\sigma}$, $m_{0\pi}$ and $2m_{0\pi}$ as a function of $T$.
One sees that in the low-temperature region, Pole {\sc II} behaves like $m_{0\sigma}$, while in the high-temperature region not Pole {\sc II} but Pole {\sc I} behaves like $m_{0\sigma}$.
We can interpret this behavior as the level crossing of $\sigma$ and $\pi\pi$ states which have the same quantum number.
The tree level masses can be identified as the energies of the states of $\pi$ and $\sigma$ without interaction between them.
When $T$ is low, $m_{0\sigma}$ is larger than twice $m_{0\pi}$.
But at some $T$ below $T_c$, they coincide with each other and beyond that temperature $m_{0\sigma}$ becomes smaller than twice $m_{0\pi}$.
Accordingly, when $T$ is low  and $m_{0\sigma} > 2m_{0\pi}$ the higher pole,  Pole {\sc II}, corresponds to $\sigma$ and the lower pole, Pole {\sc I}, to $\pi\pi$.
As $T$ increases and $m_{0\sigma} \sim 2m_{0\pi}$, the states of $\sigma$ and $\pi\pi$ largely mix because of their approximately degenerated energies and the strengthened coupling. For this reason Pole {\sc I} represents a mixture of $\sigma$ and $\pi\pi$.
As $T$ increases further and $m_{0\sigma} < 2m_{0\pi}$, the coupling of $\sigma$ and $\pi\pi$ tends to diminish and now the lower pole, Pole {\sc I}, is dominated by $\sigma$ in contrast to the case of low $T$.


In summary, we studied how we can understand the change of the spectral function and the pole location of the correlation function for sigma at finite temperature, which were previously obtained in the linear sigma model with a resummation technique called OPT.
There are two relevant poles in the sigma channel.
One pole is the original sigma pole which shows up as a broad peak at zero temperature and becomes lighter as the temperature increases.
The behavior is understood from the decreasing of the sigma condensate, which is consistent with the Brown-Rho scaling. 
The other pole changes from a virtual state to a bound state of $\pi\pi$ as the temperature increases which causes the enhancement at the $\pi\pi$ threshold.
The behavior is understood as the emergence of the $\pi\pi$ bound state due to the enhancement of the $\pi\pi$ attraction by the induced emission in medium.
The latter pole, not the former, approximately degenerates with $\pi$ above the  critical temperature of the chiral transition.
This means that the observable \lq\lq$\sigma$" changes from the former to the latter pole, which can be interpreted as the level crossing of $\sigma$ and $\pi\pi$ at finite temperature.
Although the results in the present paper are based on the one-loop calculation, the basic trend is expected to remain even after higher loop effects are taken into account.
Namely, there might be two peaks in the sigma spectrum corresponding to two relevant poles
and the lower peak, which has a width due to higher loop effects, would
degenerate with $\pi$.
It is confirmed at least in the calculation taking account of the effect 
of constant pion thermal width \cite{Hidaka:2003mm}.

\end{document}